# Experiential probabilistic assessment of cloud services

**Mahdi Fahmideh**[1], **Ghassan Beydoun**[2], **Graham Low**[3]

[1] University of Wollongong, Australia
[2] University of Technology Sydney, Australia
[3] University of New South Wales, Australia

**Abstract.** Substantial difficulties in adopting cloud services are often encountered during upgrades of existing software systems. A reliable early stage analysis can facilitate an informed decision process of moving systems to cloud platforms. It can also mitigate risks against system quality goals. Towards this, we propose an interactive goal reasoning approach which is supported by a probabilistic layer for the precise analysis of cloud migration risks to improve the reliability of risk control. The approach is illustrated using a commercial scenario of integrating a digital document processing system to Microsoft Azure cloud platform.

## 1. Introduction

There are many reports of failure of cloud adoption. Unexpected challenges have been faced by both service consumers and providers (Chow, Golle et al. 2009, Pepitone 2011, Linthicum 2012, Tsidulko 2016). Expecting service providers to guarantee SLA is not always realistic (Qiu, Zhou et al. 2013, Gill, Smith et al. 2014). Hidden risks in cloud services operations may impact quality of system goals. These include issues such as vendor lock-in, data security, or interoperability and they are costly to resolve after moving systems to the cloud. A notable example is Infoplus's report that a new high-frequency cloud-based trading system started making unprofitable trades up to 40 times per second. This forced New York Stock Exchange to halt all trading and caused a loss of over $440 million (Musiienko December 2017). Infoplus' survey conducted by iLand, found that almost 57% and 44% of Amazon Web Service (AWS) and Microsoft Azure users have reported stalled or failed cloud adoption. Another survey conducted by NTT Communication concludes that 41% of the decision makers believed that migrating complex systems to the cloud is more trouble than its worth (Communications 2015).

Complete knowledge about those hidden risks is typically not available. A system architect would be tasked with re-architecting existing legacy systems to the cloud to reduce infrastructure cost and achieve higher throughput. However, they would be unsettled with many questions such as (i) Will higher system throughput be achievable in all situations? (ii) What risks are likely to obstruct reducing infrastructure cost and high throughput? or (iii) how can such risks be negated or reduced? A deeper analysis is required to identify risks and obstacles for quality goals. Indeed, a rigorous study of their severity of consequences is a must to ensure appropriate mitigation tactics are in place. Architects require more flexibility to explore hidden threats and avoid premature decisions (Tran, Keung et al. 2011, C.Tang 2013, Pahl, Xiong et al. 2013). Assisting them in this is the focus of this paper.

The paper continues our research efforts in facilitating cloud migration. It focuses on anticipating exceptional situations that can disrupt migration goals. In earlier work, we identified key challenges in overall re-architecting legacy system architectures to the cloud (Fahmideh, Daneshgar et al. 2017). That work provided guidelines to identify both high-level technical and non-technical architectural requirements. In a more recent effort (Fahmideh and Beydoun 2018), we provided an approach to store this architectural knowledge for the purpose of its reuse. A knowledge repository was developed. The current paper operationalises the deployment of such a repository by providing a decision making layer based on a probabilistic assessment of disruptive obstacles. This decision layer supports a system architect to appropriately respond and manage risks in adoption of cloud services. The reasoning approach is iterative and facilitates a top down refinement of obstacles. The refinement and the probabilistic assessment are interleaved and run iteratively. The probability of an obstacle is estimated at each level of refinement. In the final decision support phase, all probabilities are collectively used to assess the impact of an obstacle on the system quality goals. A precise semantic is used to represent the goal satisfaction and obstacle estimation. This underpins the support and propagation rules to enable formal goal reasoning. This is a novel probabilistic foundation for assessing obstacle severity and the concomitant degree of goal satisfaction.

The rest of the paper is organised as follows. Section 2 provides a background on legacy systems migration to the cloud following and an overview of related work. Section 3 details the framework's components using an example of Amazon service adoption for a legacy system. Section 4 shows the applicability of the framework in a scenario of goal-obstacle analysis for moving a digital document processing systems to Microsoft Azure cloud platform. Section 5 summarises the paper and discusses limitations and future directions of the research.

## 2. Probabilistic goal-obstacle analysis

Goal-oriented modelling frameworks such as KAOS and i* provide means for the elicitation, elaboration, and analysis of goals, from high-level strategic goals to concrete and technical details (Yu and Mylopoulos 1994). KAOS (Keep All Objects Satisfied) supports different levels of formalism for expressing goals and reasoning about them. The levels can vary from semi-formal analysis goal models to formal when a precise reasoning is required (Dardenne, Van Lamsweerde et al. 1993, Van Lamsweerde and Letier 2004). The goals are iteratively refined through top-down approach (by asking *how* questions to refine goals into sub-goals) as well as bottom-up approach (by asking *why* questions to identify parent goals). KAOS's concepts used in this research are detailed in what follows.

**Goal**. A *goal* is a desired property or statement to be satisfied by a system through the collaboration of agents or actors. Goals may vary in abstraction from business level to fine-grained technical (Letier 2001). Linear temporal logic (LTL) may be used for formally representation of a goal. It is in a general form like $C \rightarrow \Theta T$ where $\Theta$ represents a LTL operator such as: o (next state), e (sometimes in the future), ◊ (sometimes in the future before deadline d), □ (always in the future), ◊ $\leq_d$ (always in the future up to deadline d), *W* (always in the future unless), *U* (always in the future until), and where P → Q means □ (P→Q).

**Obstacle.** An *obstacle* to a goal is an exceptional situation/condition that prevents the goal from being satisfied (Potts 1995, van Lamsweerde and Letier 2000, van Lamsweerde 2004). Obstacles can be technical or non-technical nature. Obstacles and goals are dual notions. They capture desirable and undesirable conditions, respectively (Letier 2001).

**AND/OR refinement**. In a goal model, goals are structured through AND/OR refinement mechanisms. They identify how goals contribute to each other. An AND-refinement link decomposes a parent goal into a set of fine-grain child goals where satisfying all child goals yield satisfaction of the parent goal. An AND-refinement should be complete and consistent (Darimont and Van Lamsweerde 1996). A refinement is said to be complete if all child goals suffice the satisfaction of the parent goal. This is represented by $\{SG_1, SG_2, SG_3, \ldots SG_n, Domain\} = G$ (complete refinement). If there are no contradictions between all goals, then the refinement is said to be consistent in the domain, i.e. $\{SG_1, SG_2, SG_3, \ldots SG_n, Domain\} \neq flase$ (consistent refinement). On the other hand, OR-refinement link decomposes a goal into a set of alternative ways to satisfy a top goal and it is represented as *for all i: $\{OO_i, Domain\} = O$*. Goals may be in contradiction. Conflict link is can be used for this purpose but such links are out of the scope our work at this stage.

AND/OR refinements can also be defined for obstacles. An AND-refinement means that the occurrence of the parent obstacle depends on all its child obstacles. An OR-refinement means the occurrence of a root obstacle depends on the occurrence of at least one its child sub obstacle. The completeness and consistency conditions are also the same for the obstacles. Figure 1 shows the graphical notation of this notion using during analysis.

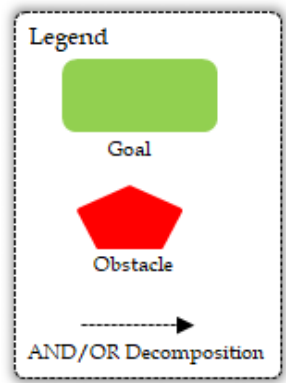

Figure 1 goal obstacle analysis diagram notation

The proposed approach uses the probabilistic view for goal and obstacle analysis grounded on a system-specific situation as suggested in (Cailliau and van Lamsweerde 2013). A goal defines a possible set of behaviour. The probability of a goal satisfaction is defined in view of probability of observing such behaviours (Cailliau and van Lamsweerde 2013). For a goal $C \rightarrow \Theta T$, the probability of satisfaction of the goal is the ratio between (i) number of possible behaviour satisfying the goal's antecedent C and consequent $\Theta T$ and (ii) number of possible behaviour satisfying condition C. If the probability of the goal satisfaction is 1, then the goal is fully satisfied. A goal might be partially satisfied due to occurrence of some obstacles. The probability of an obstacle occurrence depends on the satisfaction of its conditions. These are defined in what follows.

**Definition 1**. The probability of satisfaction of a goal in view of its possible obstructions is called estimated probability of satisfaction (EPS) (Cailliau and van Lamsweerde 2013) and is computed from the goal model. The EPS of a goal G is shown by P(G).

**Definition 2**. The minimal probability of satisfaction of a goal is called required degree of satisfaction (RDS) and is specified by existing standards and regulations in a domain of interest (Cailliau and van Lamsweerde 2013). A goal G is probabilistic if 0 < RDS (G) <1. Note that RDS value is not unique and can vary from one scenario to another. This value may be obtained from domain experts, user experience, or existing knowledge about the system.

**Definition 3.** Based on the EPS and RDS, the gap between estimated and expected probabilities can be measured. If EPS ≥ RDS, then the required goal satisfaction is reached. If EPS < RDS, then the goal is not satisfied and the gap should be investigated and reduced to the extent possible. The severity of violation (SV) from a goal G is defined as:

$$SV(G) = RDS(G) - P(G) \qquad (I)$$

The above equations are used to measure the probability of goal satisfaction by propagating probabilities from leaf obstacles towards top goals in a goal model. Hence, the estimated probability of leaf obstacles should be first provided. These estimates are then propagated up towards the root obstacles, leaf goals, and finally parent goals to compute the probabilities of goal satisfaction in view of obstacle occurrence. Cailliau et. al. define the following propagation process (Cailliau and van Lamsweerde 2013):

*(i) From leaf obstacles towards root obstacles.* The system architect should rely on domain information to estimate the probability occurrence of leaf obstacles in the refinement goal model. The information sources include the specification of cloud services, system developers or end-users' experience, consultation with domain experts, and statistical data about legacy systems which can be obtained through techniques like interview or Delphi method.

In an AND-refinement, a parent obstacle occurs if all its sub-obstacles (SO) also occur. Thus, the probability of the parent obstacle equals the probability of all sub-obstacles and their combined occurrence towards the satisfaction of the parent obstacle. This is computed as follows: $P(O) = P(SO_1) * P(SO_2) * P(SO_3) * \ldots * P(O|SO_1, SO_2, SO_3, \ldots)$ \hspace{1em} (II)

In equation (I), the architect also needs to know from the domain information how often the occurrence of leaf obstacles O1, O2, O3, and … causes the parent obstacle O happens. For an OR-refinement, the probability of the parent obstacle *not* occurring which equals the probability that none of the children obstacles yield a satisfaction of the parent obstacle. This is computed as follows:

$$P(O) = 1 - (1 - P(SO_1) * P(O|SO_1)) * (1 - P(SO_2) * P(O|SO_2)) * \ldots) \hspace{1em} (III)$$

(ii) *From root obstacles towards leaf goals.* The probability of not satisfying of a leaf goal (LG) is given by the probability that the root obstacle occurs (RO) and such occurrence results in not satisfying of the leaf goal. This is presented using the following equation:

$$1 - P(LG) = P(RO) * P(\neg LG|RO) \hspace{1em} (IV)$$

If a leaf goal is obstructed by multiple obstacles, then the goals is satisfied when any obstacles occurs. This is computed as follows:

$$P(LG) = (1 - P(O_1) * P(\neg LG|O_1)) * (1 - P(O_2) * P(\neg LG|O_2)) * \ldots \hspace{1em} (V)$$

(iii) *From leaf goals towards root goals.* The satisfaction probability of a parent goal depends on probabilities of satisfaction of its leaf goals. Thus, the reduced degree of satisfaction of an obstructed leaf goal should be propagated upwards in the goal model in order to specify consequences of all obstacles. For example, a parent goal with two leaf goals is satisfied if all the leaf goals are satisfied, or satisfaction of the first goal is sufficient to satisfy the parent goal, or the satisfaction of the second one is sufficient to satisfy the first one (Cailliau and van Lamsweerde 2013). To specify the consequence of the obstacles to the quality goals, the computed probabilities for all leaf obstacles are propagated upwards towards goals. This enables computing EPS of higher-level goals in view of its possible obstacles deviating from RDS. A parent goal is satisfied if its sub goals are satisfied. Hence, we have:

$P(G) = P(SG_1, SG_2) * P(G|SG_1, SG_2) + P(SG_1, \neg SG_2) * P(G|SG_1, \neg SG_2) +$
$P(SG_2, \neg SG_1) * P(G|SG_2, \neg SG_1) + P(\neg G|SG_1, \neg SG_2) * P(G|\neg SG_1, \neg SG_2)$ \hspace{1em} (VI)

## 3. Proposed approach

The approach is a four-step process. It starts with elicitation of high-level goals for the cloud adoption to empower legacy systems. It ends with a set of critical obstacles obstructing goal satisfactions. The output is a list of critical obstacles that need to be dealt with. To illustrate, we use an example scenario of moving the data storage of a legacy system to Amazon Simple

Storage Service (S3) so that end users can get the content directly from Amazon S3. S3 provides a secure, durable, highly-scalable cloud storage to store and retrieve any amount of data from anywhere on the web (AmazonS3). As earlier mentioned, the knowledge repository has collections of goals, obstacles, and resolution tactics which will be used to illustrate the goal analysis. The repository itself was developed in (Fahmideh and Beydoun 2018). The steps required are detailed in what follows.

*Step 1. Specifying goals*

The system architect first identifies goals that are expected to be satisfied by integrating the system with cloud services. In the running scenario, deploying the legacy system database on Amazon S3 are expected to positively contributes towards the following five goals: Achieve [Reduced IT cost], Achieve [Improved response time], Achieve [Improved availability], and Achieve [Improved consistency]. For example, the goal Achieve [Improved response time] has the following specification.

> **Goal** Achieve [Improved response time]
> **Category** Performance Goal
> **Definition** [Using Amazon S3, transferring 100 terabyte live data stream with internet connection 1000 Mbps at 80% network utilization from local network should not take no more than one week].
> **Formal spec** $\forall\, d: data, (d.submitted \rightarrow \Diamond \leq_{7\ days} d.processed\,)$
> **RDS 95%**.

This goal is satisfied if its two child goals are satisfied. In other words, AND-refinement for the parent goal Achieve [Improved response time] captures a combination of two leaf goals Achieve [Reduced data uploading time] and Achieve [Reduced query processing time] entailing the parent goal should be completely satisfied (Figure 2). The specification of the goal Achieve [Reduced data uploading time] is:

> **Goal** Achieve [Reduced data uploading time]
> **Category** Performance Goal
> **Definition** [Using Amazon S3, transferring 1 terabyte live data with internet connection 1000 Mbps at 80% network utilization from local network should not take no more than 24 hours].
> **Formal spec** $\forall\, d: data, (d.submitted \rightarrow \Diamond \leq_{24\ hours} d.processed\,)$
> **RDS 90%**.

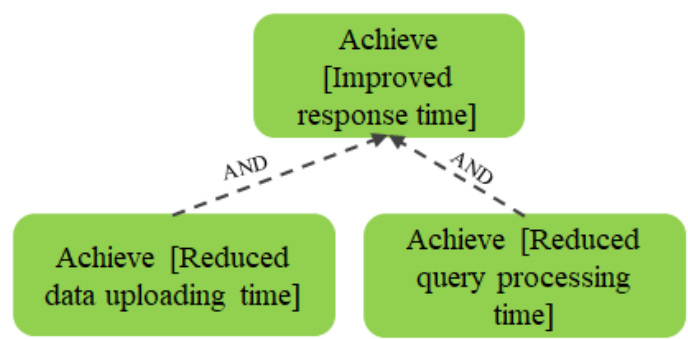

Figure 2 Refining the parent goal Achieve [Improved response time] to child goals

*Step 2. Identifying obstacles*

For each quality goal, the system architect explores potential obstacles. For each quality goal, they shortlist those probable ones (Figure 3). This identification of obstacles is based on information sources such as developers, user experience, statistical data, and available technical accounts about Amazon S3. In this scenario, former developers' experience concedes that the leaf goal Achieve [Reduced data uploading time] is likely hampered by two obstacles *Performance variability of Amazon S3* and *Geographical distance*. In addition, for the purpose of message processing, the current legacy system uses Kafka technology, i.e. a common open source technology for massive scale publishing and subscribing message queues. Integrating this technology with Amazon S3 storage may cause latency in maintaining data consistency. From past experience in using Amazon S3, there is no guarantee of the exact time to upload the system data to the S3 servers due to their workload unpredictability. The uploaded data by a user may be stored in S3 data storage for an extended period of time which may cause a data inconsistency issue. This obstacle, named here as *Latency for moving data from Kafka to S3*, is against the goal Achieve [Improved consistency]. As shown in Figure 3, other leaf obstacles that are modelled by the system architect are *Service transient fault*, *S3 outage*, *Department downsizing*, and *Extra management effort per annum*. The obstacles *Extra cost of training new data integrator* and *Extra cost for monitoring tools* are refinements of obstacle *Extra management effort per annum*. The obstacles *High uploading time for blob* and *Low throughput to write bucket* are refinements of *Performance variability of Amazon S3*. *Local electrical storm*, *S3 power outage*, *S3 data centre outage*, *I/O issues of servers*, and *Local network disruption* are refinements of the obstacle *Service outage*. No obstacle is identified for the goal Achieve [Reduced query processing time]. Figure 4 shows all refinements to root obstacles. AND/OR refinements are used to show sub obstacles of a root obstacle. AND-refinements include a combination of sub obstacles that aggregately cause the occurrence of a parent obstacle. OR-refinements represent a set of alternative sub obstacles where the occurrence of each will cause the parent obstacle.

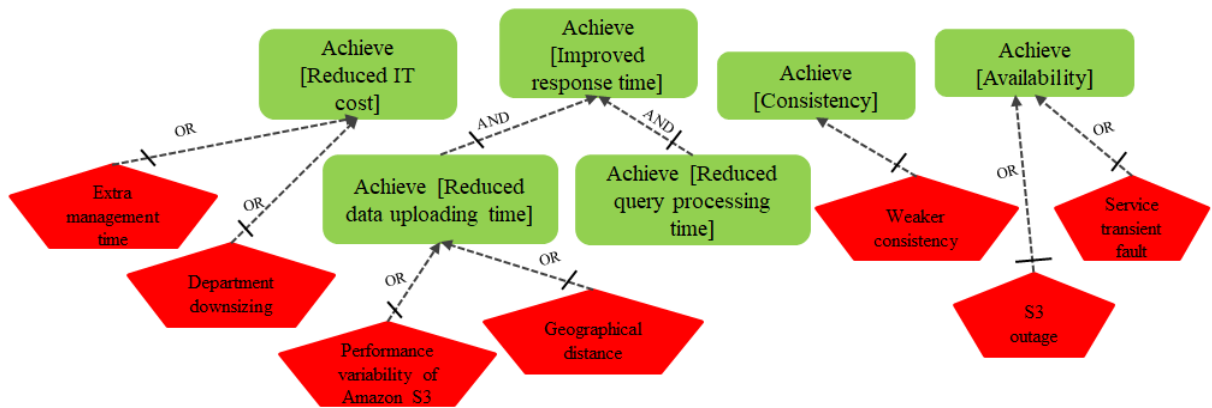

Figure 3. Identified obstacles against achieving quality goals Achieve [Reduced IT cost], Achieve [Reduced response time], Achieve [Improved consistency], and Achieve [Improved availability]

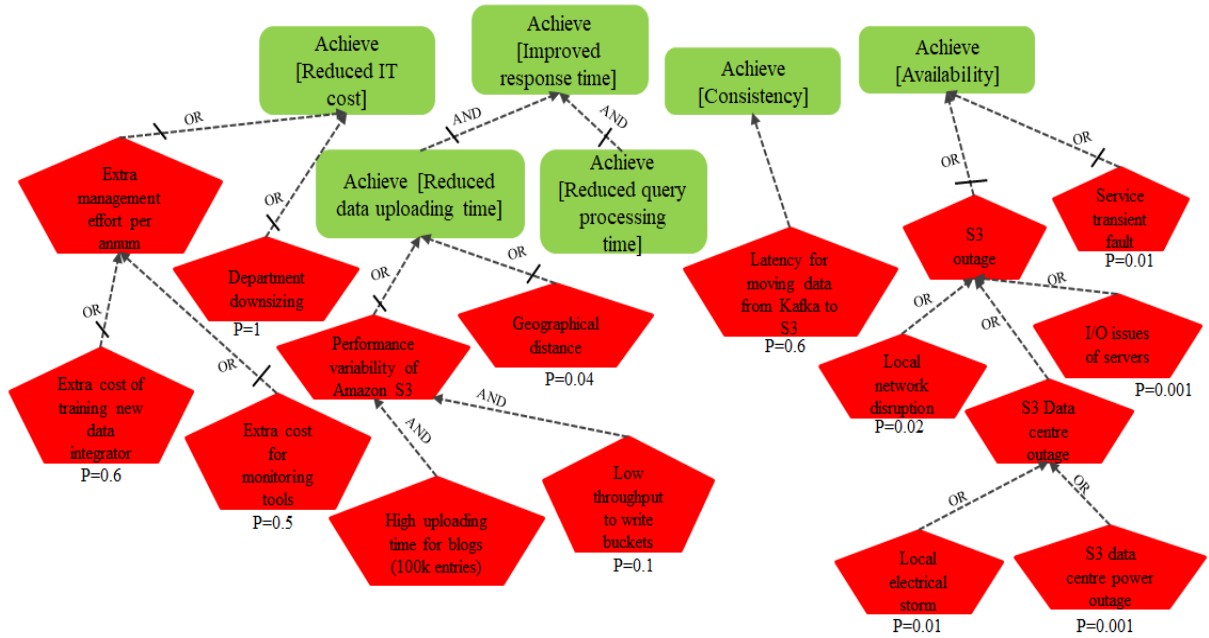

Figure 4. Refining the root obstacles to leaf obstacles

### Step 3. Assessing probability of obstacles

Table 1 shows a sample of 12 leaf obstacle estimations based on the developers' opinions, and existing statistics about Amazon S3 server, memory and I/O usage, and legacy system performance. These estimates can be refined as the goal analysis proceeds.

Table 1 Estimated probabilities for the leaf obstacles based on domain information

| Obstacle | Definition | Probability |
|---|---|---|
| Department downsizing | Moving a legacy system database to S3 definitely causes change in roles and responsibilities defined in the maintenance team of IT department. | 1 |
| Extra cost of training new data integrator | Although developers have expertise in using legacy-based tools for data integration, they require training in using new data integration tools specific for combining and import/export legacy data and S3. | 0.6 |
| Extra cost for monitoring tools | New tools should be installed and a new role should be appointed for monitoring the database performance in S3. | 0.5 |
| Geographical distance | It is likely that buckets stored in S3 servers are located in distance far from the local network of the company which may cause high uploading time. | 0.04 |
| High uploading time for blobs (100k entries) | Measurements of storing ten set of 100k buckets in S3, over one week period, showed that transferring one of the buckets took more than 24 hours. | 0.2 |
| Low throughput to write buckets | In high workload, storing one set of 100k buckets in the database caused excessive delay. | 0.1 |
| Latency for moving data from Kafka to S3 | In 60% of cases, there is high latency for off-line moving data from Kafka to S3 storage. Due to server workload, it is highly probable that end users may observe stale and out of date data. | 0.6 |
| Local electrical storm | The company is located in a rainy geographical area and has recently experienced electrical storm damaged power equipment at one of the data centres. | 0.01 |
| S3 data centre power outage | Drop time of Amazon S3 servers is unlikely to occur. | 0.001 |
| Local network disruption | Network misconfigurations or the level of workload may cause a local network disruption. | 0.02 |
| I/O issues of servers | I/O issues may occur but with very low probability. | 0.001 |
| Transient fault of service | Developers' experience shows the probability of transient faults when legacy system tries to connect cloud services. | 0.01 |

The probabilities of the occurrence of the obstacles in the goal model are now computed starting from lead obstacles. Depending on the structure of the goal model, this has tree possible steps (Cailliau and van Lamsweerde 2013):

*(i) From leaf obstacles towards root obstacles.* As noted in Section 2, in an AND-refinement, a parent obstacle occurs if all its leaf obstacles occur. Similarly, in OR-refinement, a parent obstacle occurs in the case of occurrence of any leaf obstacles. Given the probabilities of leaf obstacles in Table 2, the propagation rules for AND/OR-refinements in equations (II) and (III) are applied bottom-up to compute probabilities for the parent obstacles *Extra management effort per annum*, *performance variability of Amazon S3*, and *S3 data centre outage*.

With respect to the obstacle *performance variability of Amazon S3*, the estimations for the leaf obstacles are namely 20% of data transfer from local network to Amazon S3 experiences a high uploading time, 4% uploading data are performed on servers which are too far from the local network. From the domain information, it is known that 95% of delays in data uploading causing low throughput in writing buckets is due to the performance variability of Amazon S3. Given that, the propagation rule for the AND-refinement results in the probability of the root obstacle *Performance variability of Amazon S3*:

> P (performance variability of Amazon S3) = P (High uploading time for blogs) * P (Low throughput to write buckets) * (performance variability of Amazon S3 | High uploading time for blogs, Low throughput to write buckets) = 0.2 * 0.1 *0.95 = 0.019

This means that the performance variability of S3 for the data processing occurs in almost 2 % of cases. The propagation rule for OR-refinement is used for the root obstacle *Extra management effort per annum*:

> P (Extra management effort per annum) = 1 – P (Extra cost of training new data integrator) * P (Extra cost for monitoring tools) = 1– (1–0.5*0.99) * (1–0.6*0.99) = 0.205

In above the proportion of both obstacles is considered 99%. Similarly, the probability of *S3 data centre power outage* is computed using equation III as follows:

> P (S3 data centre outage) = 1– (1–P (Local electrical storm) * (P (S3 data centre power outage | Local electrical storm) * (1-P (S3 power outage) * P (S3 data centre power outage | S3 power outage)) = 1 – (1–0.01*0.99) * (1–0.001*0.98) = 0.010

According to the statistical data, there is a 2% probability of *local network disruption*, 0.01% of *I/O issues of servers*, 0.1% *S3 data centre power outage*, 1% *Local electrical storm*, and 1% *S3 power outage*. The proportion of *local network disruption*, *I/O issues of servers*, and *S3 data centre power outage* is respectively 99%, 98%, and 100%. The propagation rule for OR-refinement in equation (III) yields the following probability for the parent obstacle *S3 outage*:

> P (S3 outage) = 1 – (1–0.02*0.99) * (1-0.001*0.98) * (1-0.01*1) = 0.03

The above value means the S3 outage occurs in %3 of cases.

(ii) *From root obstacles towards leaf goals*. Back to Figure 4, the goal Achieve [Reduced IT cost] is satisfied when none of the leaf obstacle occurs. The probability of satisfaction for this goal is computed using equation V which is:

> P (Achieve [Reduced IT cost]) = (1 – P (Extra management effort per annum) * P (¬Reduced IT cost | Extra management effort per annum) * (1–P (Department downsizing) * P (¬Reduced IT cost | Department downsizing) = (1 – 0.6*0.6)*(1–1*1) = 0

This means that using Amazon S3 will not certainly reduce the IT cost unlike the initial expectation. In addition, the probability of satisfaction for the leaf goal Achieve [Reduced data uploading time] is computed using equation IV which is:

P (Achieve [Reduced data uploading time]) = (1 – P (Performance variability of Amazon S3) * P (¬Reduced data uploading time | Performance variability of Amazon S3) * (1 – P (Geographical distance) * P (¬Reduced data uploading time | Geographical distance) = 0.9

The above value means in 90% of cases, 1 terabyte live data with internet connection 1000 Mbps at 80% network utilization, will be transferred from the local network to S3 within the prescribed 24 hours. Furthermore, using the same equation, the probability of the goal Achieve [Availability] is:

P (Achieve [Availability]) = (1 – P (S3 outage) * P (¬Availability | S3 outage) * (1 – P (Service transient fault) * P (¬Availability | Service transient fault) = 0.9

(iii) *From leaf goals towards root goals*. In the exemplar model (Figure 4), the system architect checks if the model satisfies the expected threshold which is the response time for processing of requests, i.e. goal Achieve [Improved runtime response], and this should be satisfied in 90% of cases (RDS=0.9). For the leaf goal Achieve [Reduced data uploading time] the computed satisfaction probability is 0.9. Based on the developers' experience, there is no obstacle against the goal Achieve [Reduced query processing time]. Thus, the probability of satisfaction of the goal Achieve [Improved response time], which is an AND-refinement link, is 1*0.9 = 0.9. The resulting EPS for this goal is 90% which means the adopting Amazon S3, as modelled, is not able to satisfy the expected standard 95%.

*Step 4. Identifying critical obstacles*

The system architect is particularly interested in identifying obstacles that cause a sever violation. Hence, only one single leaf obstacle is considered and the probabilities for the rest are set to 0. Propagation values towards the root goal is performed, and a violation severity for the root goal is computed.

Whilst some leaf obstacles may have small probabilities, they may still be more important than others (Cailliau and van Lamsweerde 2013). It is also important to realise that in the case of generating many leaf obstacles, the computation of obstacle consequences on leaf goals and root goals may be difficult. The identification of critical obstacles is a multi-criteria optimisation problem where the aim is to find the minimal set of leaf obstacles that maximise the violation severity of high-priority goals. It has three steps: (i) generating all leaf obstacle combinations, (ii) computing SV(G) for each obstructed goal (if required, goals are weighted), and (iii) sorting leaf obstacles combination based on their severity. Identification and prioritisation of critical obstacles, not our focus, can be performed using common techniques covered elsewhere such as AHP and brute-force technique. The critical obstacles should be tackled prior to the enactment of cloud services. The obstacle representation from (Fahmideh & Beydoun, 2018) is used. A catalogue of resolution tactics is stored in the repository. These are used by the system architect to deal with any critical obstacles. An excerpt of resolution tactics which can be used during obstacle handling is shown in Table 2. For example, the system architect can adopt resolution tactics *Encrypt data* to handle *obstacles Data disclosure*, *Session hijacking*, and *Insecure data location*.

Table 2. An excerpt of resolution tactics for tacking obstacles

| Obstacle(s) | Resolution tactic | Definition |
|---|---|---|
| Incompatible pluggable cloud services, Incomplete APIs, Incompatible data types, Operating system incompatibility, Machine-image incompatibility, | Develop adaptor/wrapper | Add adaptors for resolving mismatches, occurring at runtime system execution, between legacy system components and cloud services. |
| Virtual machine contextualization incompatibility, API incompatibility across multiple cloud, Proprietary APIs | | |
| Tight dependencies | Decouple system components | Decouple the legacy system components from each other. Use mediator and synchronisation mechanisms to manage interaction between the loosely coupled components in the cloud environment. |
| Message passing, Data disclosure | Encrypt/decrypt message passing | Add support for the runtime encryption/decryption of message transition between components in the on premise network and cloud environment. |
| Code disruption, System source codes propriety, Data disclosure | Obfuscate code | Protect unauthorised access to code blocks of components by other tenants that are running on the same cloud provider. Use encryption mechanisms in the sense that no other tenants will be able to access, read, or alter the code blocks with the components when running in the cloud. |
| Tenant interfere, Data disclosure | Isolate tenant | Enable multi-tenancy in the system. Based on multi-tenancy requirement (i) define tenant-based identification and hierarchical access control for tenants and (ii) separate tenant data using authorization and authentication mechanisms. |
| Message passing | Tune message granularity | Define suitable granularity for messages, which are passing between components hosted on local network and the cloud, based on the degree of functionality that is offered to the service consumer and consumer's infrastructure capability to process the messages. A proper message granularity can be identified or predicted based on pieces of data actually used by system or using heuristic functions to understand the number of interaction between system components over the cloud network. |
| Incompatible data types, Incompatible data operations | Adapt data | Convert legacy data types to the data type of target cloud database solution. Also, add an extension component to the legacy which includes a set of commands to be performed by the legacy or cloud. The emulator supports missed database functionalities of cloud database solution provider. |
| Department downsizing, Resistance to change | Involve staff with cloud adoption process | Involve staff and stakeholders actively in the cloud adoption process and give them insight of benefits of the cloud and organisational change. |
| Tenant interfere | Define an authorization | Add a component determining if a tenant has privilege to perform a given action over the database. |
| Data remanence, Data interruption, Data disclosure, Session hijacking, Insecure data location | Encrypt data | Use data encryption mechanisms prior outsourcing or hosting system data to the cloud. |
| Tenant interfere, Data interruption | Filter unauthorised requests | Add support to filter unauthorized data access received from users at the edge of premise or cloud network as early as possible to avoid unauthorized network traffic. |
| Performance variability of cloud service | Use multiple cloud servers | Deploy and replicate system components in several clouds. |
| Scaling latency, Low middleware performance, Service latency | Add intermediation | Implement an intermediate layer (mediator components) between legacy system and cloud |

| | | services that decouple legacy systems from cloud specific APIs. This helps to create intermediate APIs and get indirect service from the cloud. |
|---|---|---|
| State-based dependency | Make system stateless | Provide support in the system to the handle safety and traceability of tenant's session when various system instances are hosted in the cloud. |
| Extra testing effort | Prioritize tests | Perform test cases on the basis of their importance and criticality. |
| Licensing issue | Resolve licensing issue | There alternative sub-tactics: (i) negotiate with system owner to make a suitable licensing model which satisfies all parties, (ii) extend legacy system with a new component (e.g. VPN tunnel) in a way that cloud services can be indirectly offered to them, and (iii) enable a license tracking mechanism through monitoring connections between the software system and cloud resources. |
| Session hijacking | Update patches | Perform regular patch update across system components in the cloud. |

## 4. Validation

The case study used for the validation is based on a scenario of moving a Web-based Digital Document Processing (DDP) legacy system to the cloud (Rabetski 2012, Rabetski and Schneider 2013). *InformIT* is a small independent software vendor in Sweden. It has developed the DDP to offer services to medium and large organisations with adequate infrastructure and technicians. *InformIT* planned to further expand the system's services to also support small companies that could not afford the current required financial commitment to use the system. Deploying the DDP in the cloud would enable small companies to utilise its services without purchasing the infrastructure. The system architect was interested in analysing obstacles that such a transition would experience and accordingly handle them beforehand. The goal modelling steps were performed as described in the following.

*Step 1. Specifying goals.* The system architect identified four goals Achieve [Performance], Achieve [Integrity], Achieve [Portability], and Achieve [Accountability] to be satisfied by the moving DDP to Microsoft Azure cloud platform. These goals are defined as follows:

> **Goal** Achieve [Security]
> **Category** Security goal
> **Definition** [DDP's documents should not be accessible/readable by other tenants that are running on same Azure servers].
> **Formal spec** ($\forall\ doc: DDP's\ document\ \rightarrow \Box doc.processed\ by\ registered\ tenant$)
> **RDS: 100%**
>
> **Goal** Achieve [Performance]
> **Category** Performance goal
> **Definition** [acceptable system throughput for rendering a digital document should be no more than 4.9 seconds].
> **Formal spec** ($\forall\ doc: document,\ doc.submitted\ \rightarrow \Diamond \leq_{4.9\ seconds} doc.processed$)
> **RDS: 95%**
>
> **Goal** Achieve [Testability]
> **Category** Testing goal
> **Definition** [the whole process of performing various tests scripts of system components should be doable within a specific time limit].
> **Formal spec** ($\forall\ com: DDP's\ component\ \rightarrow \Diamond \leq_{1\ working\ day} com.tested$)
> **RDS: 95%**
>
> **Goal** Achieve [Integrity]

**Category** Integrity goal
**Definition** [system components should be able to invoke cloud services].
**Formal spec** ($\forall\ com: DDP's\ component\ \rightarrow \Diamond\ com.invoked.cloud\ services$ )
**RDS: 95%**

**Goal** Achieve [Portability]
**Category** Portability goal
**Definition** [DDP's documents should be readable and processable in both platforms].
**Formal spec** ($\forall\ doc: DDP's\ document\ \rightarrow \Diamond\ doc.processed$ )
**RDS: 95%**

*Step 2. Identifying obstacles.* For each goal, the system architect first refined the top goals towards root obstacles, and subsequently the leaf ones that may obstruct quality goals. Information provided by developers and end users of DDP were the main source used to check if an obstacle was likely to occur or not. For example, consider the goal Achieve [Integrity]: The current DDP's APIs were not be compatible with their counterparts in the Microsoft Azure cloud platform. Thus, the leaf goal is obstructed by the root obstacles *Incompatible APIs* (i.e. Legacy's APIs and Microsoft Azure) and *Incompatibility of legacy data storage and cloud*. Furthermore, the parent obstacle *Incompatibility of legacy data storage and cloud* was also refined into two leaf obstacles. Both of these two were domain specific instantiations of the obstacle *Incompatibility of legacy data storage and cloud*. The definition of the leaf obstacles against the goal Achieve [Integrity] are depicted in Figure 5 and defined as follows:

**Obstacle** *Incompatible APIs*
**Definition** [DDP uses API's offered by .NET 2.0 and Visual Studio 2005 which may not be compatible with Microsoft Azure platforms].
**Formal spec** $\Diamond\ (\neg DDPAPIsIncompatibility$ )

**Obstacle** *Incompatible datatypes*
**Definition** [DDP datatypes are based on SQL Server Database .NET 2.0 platform which might not be compatible with Microsoft Azure database solution].
**Formal spec** $\Diamond\ (\neg LDDPDatastorageCompatibility$ )

**Obstacle** *Incompatible data operations*
**Definition** [The data operations supported by SQL Server Database .NET 2.0 platform might not be compatible with Microsoft Azure database solution].
**Formal spec** $\Diamond\ (\neg DDPDataOperationCompatibility$ )

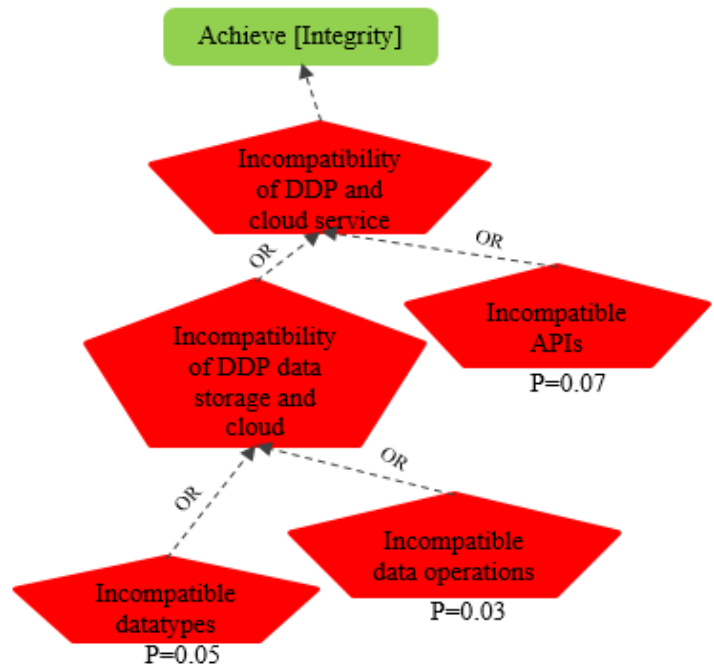

Figure 5 leaf obstacles against goal Achieve [Integrity]

Confirmed by the developers past experience, the goal Achieve [Performance] could be obstructed by the performance variability of Microsoft Azure servers. This would be out of the control of developers. This was specified by the obstacle *Microsoft Azure Middleware latency* in the goal model. This parent obstacle was also refined into three leaf obstacles *Microsoft Azure database middleware latency*, *Microsoft Azure message middleware latency*, and *Microsoft Azure transaction middleware latency*. Other relevant and potential obstacles against the goal Achieve [Performance] were *Distance from Microsoft Azure servers*, *Microsoft Azure transaction middleware latency*, and *On-premise hardware latency* (Figure 6). The refinements are shown in Figure 7.

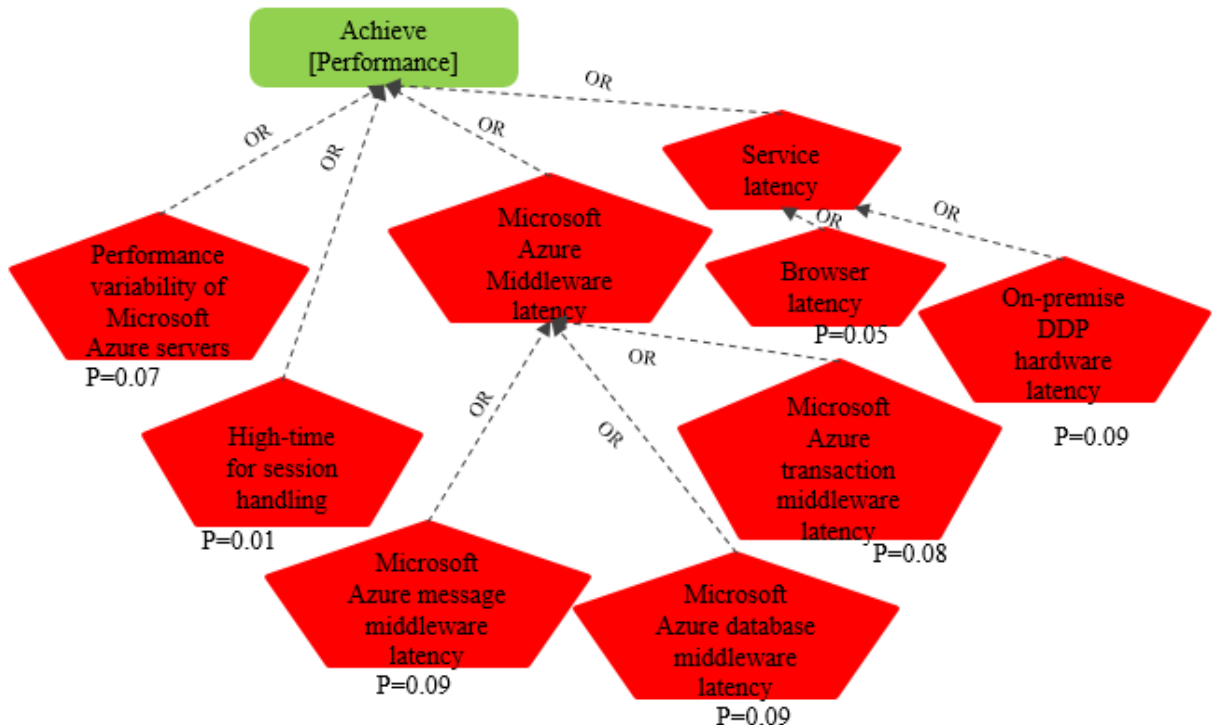

Figure 6 leaf obstacles against goal Achieve [Performance]

In total eighteen leaf obstacles that were identified in this step (Figure 5, 6, and 7) that were annotated with the estimation of their probabilities. Table 2 shows the leaf obstacle estimates. The estimates were based on the architect's judgement and consultation with the developers, and observations from the DDP past performance. All estimates can still be later refined once real statistical data become available.

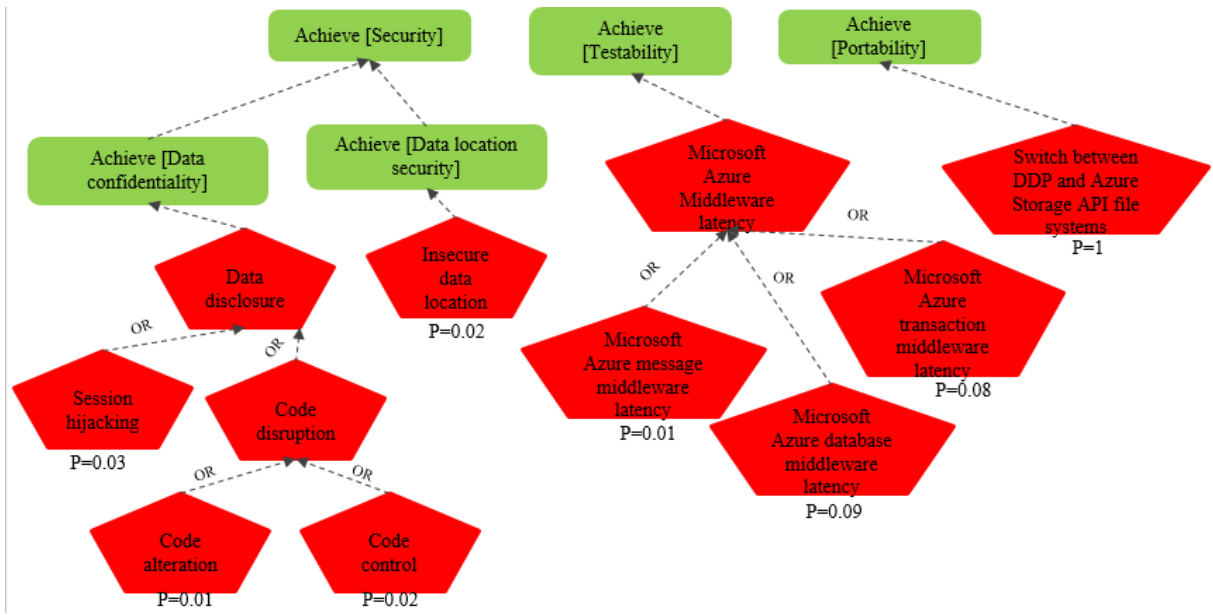

Figure 7 leaf obstacles against goals Achieve [Security], Achieve [Testability], Achieve [Portability]

Table 2 estimated probabilities for the leaf obstacles

| Leaf obstacle | Probability (%) |
|---|---|
| Switch between DDP and Azure Storage API file systems | 1 |
| Incompatible APIs | 0.72 |
| Incompatible datatypes | 0.5 |
| Incompatible data operations | 0.3 |
| Performance variability of Microsoft Azure servers | 0.07 |
| High-time for session handling | 0.01 |
| Microsoft Azure message middleware latency | 0.09 |
| Microsoft Azure database middleware latency | 0.09 |
| Microsoft Azure transaction middleware latency | 0.08 |
| Distance from Microsoft Azure servers | 0.05 |
| Browser latency | 0.05 |
| On-premise DDP hardware latency | 0.09 |
| Session hijacking | 0.03 |
| Code alteration | 0.01 |
| Code control | 0.02 |
| Insecure data location | 0.001 |

*Step 3. Assessing Obstacles.* The various probabilities were used to compute the impact of leaf obstacles on top goals via the propagation equations as described in the following.

To compute the probability of obstacle *Incompatibility of DDP data storage and cloud*, the probability of its two child obstacles should be calculated first. Given the probability estimation for the leaf obstacles *Incompatible datatypes* and *Incompatible data operations* from Table 2 and considering the proportion for having incompatible datatypes or data operations causing incompatibilities between DDP and S3 is, respectively, 0.99% and 98%, the probability of the parent obstacle *Incompatibility of DDP data storage and cloud* is obtained using equation III, producing a value of 0.65.

The probability of the obstacle *Incompatible APIs* is 0.72. The proportion *Incompatible APIs* and *Incompatibility of DDP data storage and cloud* are 0.99% and 98%. Thus, the probability of occurrence for the parent obstacle *Incompatibility of DDP and cloud service* can be again computed using equation III producing a value of 0.89. This is propagated upwards. Thus, the resulting EPS for the top goal Achieve [Integrity] is: EPS (Achieve [Integrity]) = 1 – 0.89 = 0.11

RDS for the goal Achieve [Integrity] is 1 meaning that DDP's components should be fully (100%) integratable with cloud services without occurring incompatibilities. Using equation I, SV for the goal Achieve [Integrity] is 0.95 – 0.11 = 0.84. This means the leaf obstacles are critical and should be resolved.

Similar computations are performed for the goal Achieve [Performance]. Firstly, through the OR-refinement, the root obstacle *Microsoft Azure Middleware latency* is refined into three domain-specific sub-obstacles namely *Microsoft Azure database middleware latency*, *Microsoft Azure message middleware latency*, and *Microsoft Azure transaction middleware latency*. Using equation III, the probability of the root obstacle *Microsoft Azure Middleware latency* is 0.17. The obstacle *Service latency* is refined into *Distance from Microsoft Azure servers, On-premise hardware latency, and Browser latency*. Again, using III, the computed root obstacle *Network latency* is 0.17. Using equation V, the probability of goal satisfaction for Achieve [Performance] is 0.36.

RDS for the goal Achieve [Performance] is 0.95. But SV for the goal Achieve [Performance] = 0.95 – 0.36 = is 0.6 meaning that moving the DDP to Microsoft Azure cloud platform does not render digital documents within time limit 4.9 seconds provide in 60% of cases. Countermeasure should be taken into account to satisfy the goal Achieve [Performance].

Furthermore, the satisfaction of the goals Achieve [Testability] and Achieve [Portability] depend on both the obstacles *Microsoft Azure Middleware latency* and *Switch between regular file system API to Microsoft Azure Storage API* respectively. Thus:

   EPS (Achieve [Testability]) = 1- 0.17 = 0.83

   EPS (Achieve [Portability]) = 1- 1 = 0

These values are far from RDS 1 prescribed for these goals. The system architect thus should carefully investigate these critical obstacles. The probability of satisfaction of the goal Achieve [Security] depends on the impact of the probability of leaf obstacles on the leaf goals. Given the probability estimates for the leaf obstacles in Table 2 (shown in Figure 5), the probability of the parent obstacle *Code disruption* is computed using the rules described in Section.

   P (Code disruption) = 1 – (1-0.01*0.99)*(1-0.02*0.99) = 0.02

The above is then used to compute the probabilities for the root obstacle *Data disclosure*:

   P (Data disclosure) = 1 – (1-0.02*0.99)*(1-0.03*0.99) = 0.04

The above is then used to compute the probability of the corresponding obstructed leaf goal *Achieve [Data confidentiality]*: P (Achieve [Data confidentiality]) = 1- 0.04 = 0.96

On the other hand, the probability of the goal Achieve [Data location security] is:

   EPS (Data location security) = 1- P (Insecure data location) = 1 – 0.001 = 0.99

To fully determine consequences of all obstacles, all results are propagated from leaf goals Achieve [Data confidentiality] and Achieve [Data location security] towards the root goal Achieve [Security]. Thus for the goal Achieve [Security], the following goal satisfaction probability using equation VI is computed:

   EPS (Achieve [Security]) = 0.96*0.98 = 0.94

The above means that the probability of satisfying DDP security when running on Microsoft Azure cloud platform is about 0.94. This is less than the expected RDS prescribed in the definition of the goal Achieve [Security]. The partial model in Figure 8 shows the overall satisfaction of the goals.

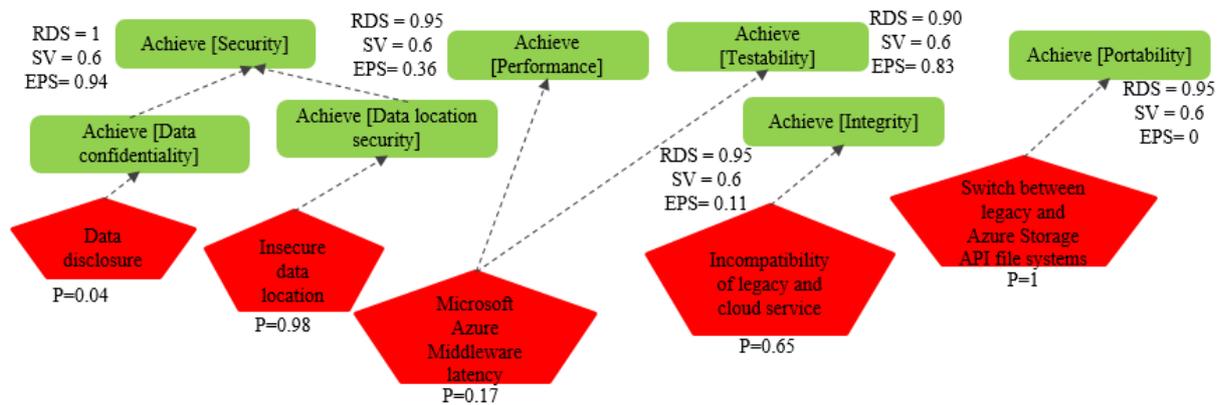

Figure 8 leaf obstacles against the root goals

*Step 4. Identifying critical obstacles*. Eleven leaf obstacles are deemed severe (as shown in Table 3). For example, *Switch between DDP and Azure Storage API file systems* with estimated probability 1, causes a violation severity of 0.95 for the goal Achieve [Portability]. This indicates that incompatibility between DDP and cloud platforms is inevitable. The leaf obstacle *Microsoft Azure database middleware latency* with the estimated probability of 0.09 causes a violation severity 0.04 for the root goal Achieve [Performance]. Therefore, in the next step of risk management the architect should define mechanisms in the new cloud-based architecture to reduce the gaps between RDS and EPS as much as possible.

As mentioned earlier, few leaf obstacles may have small probabilities but are actually more critical than others. For instance, the leaf obstacle *Code alteration* with a probability of 0.01 may obstruct the leaf goal Achieve [Data confidentiality] and subsequently the parent goal Achieve [Security]. Although its estimated probability is low, the obstacle might still be critical and might produce an overall negative impact on the security of DDP.

Table 3. Critical obstacles against goals

| Leaf obstacle | SV for root goal, i.e. (RDS – EPS) |
|---|---|
| Incompatible APIs | 0.02 |
| Performance variability of Microsoft Azure servers | 0.02 |
| Microsoft Azure message middleware latency | 0.04 |
| Microsoft Azure database middleware latency | 0.04 |
| Microsoft Azure transaction middleware latency | 0.03 |
| On-premise DDP hardware latency | 0.04 |
| Session hijacking | 0.03 |
| Code alteration | 0.01 |
| Code control | 0.02 |
| Insecure data location | 0.02 |
| Switch between DDP and Azure Storage API file systems | 0.95 |

## 5. Related work

This research bridges requirements and risk analysis of cloud services. Numerous works have looked at this. In what follows, we review some of them, in order of least to most relevance.

Less relevant works deal with risk analysis at a very high abstract level. They provide insights to system architects on system integration issues with cloud services. However, they do not provide an operational solution to examine the risk likelihood and severity during the design phase. For example, Heiser et. al. report an analysis of unique attributes of obstacles related to security goals such as data integrity and segregation. The authors found that factors such as the location independence of service provider subcontracting may endanger system security (Heiser and Nicolett 2008).

More relevant works to ours offer solutions for dealing with migration risks. These include studies that focus on risk analysis of security. e.g. Chen et. al. developed a framework for the automatic detection of conflicts and inconsistencies in user requirements and organisational policies (Chen, Yan et al. 2012). This framework suggests how cloud services may satisfy these requirements and policies. The concomitant prototype further shows how managing heterogeneous cloud infrastructure services can be undertaken for large organisations. Martens et. al., developed a quantitative model to balance costs and risk factors for outsourcing decisions regarding cloud service adoption (Martens and Teuteberg 2012). Mouratidis et. al. incorporated a modelling language along with a structured process to identify security and privacy requirements to select suitable cloud providers based on the satisfiability of the service provider (Mouratidis, Islam et al. 2013). Shirvani et. al., present a framework to support adding a module to log security information and then quantify the cloud security risks (Hosseini Shirvani, Rahmani et al. 2018).

Our formalism to represent goals complements the above efforts. The use of AND/OR refinement mechanisms enables more in-depth analysis of security risk. A clear advantage of our approach is its wide applicability to assess other risk management goals. Perhaps still closer work to our research is Saripalli et. al. (Saripalli and Walters 2010). That work presents a probabilistic approach, QUIRC (Quantitative Impact and Risk Assessment), for analysing and assessing typical security attacks to cloud based systems. To assess the impact of risks on system security, a combination of the probability of security threats and their severity is also computed. Data available from SANS (System Administration, Networking, and Security Institute) is used to facilitate the computation. One key advantage of our approach over QUIRC is our in-depth refinements of obstacles. The collective impact of lower level risks is used to assess overall risk. This enables architects to better understand the ensuing risk in the context of a multitude of obstacles. In addition, our approach is interactive and participatory. It relies on domain information to estimate risk probabilities rather than merely using public data as the main source for estimations.

Goal-oriented approach has also been used elsewhere e.g. in (Islam, Mouratidis et al. 2013). The authors use it to analyse security and privacy risks of cloud services relying on sources such as data, service/application, technical, and organisational measures. But they treat the overall risk as a single security goal, without detailed refinement and elaboration. Zardari et. al. also use a goal-oriented approach and employ Analytical Hierarchy Process (AHP) to prioritise obstacles (Zardari, Bahsoon et al. 2014). In their context, our Step 4 can be viewed as complementary to perform obstacle prioritization. Our approach here is unique. Our own earlier work presented in (Fahmideh and Beydoun 2018) highlights the need for an obstacle analysis and reusing empirical knowledge of architecture design. But the proposed approach in this research actually provides a systematic operationalisation. The work provides a probabilistic foundation for measuring the satisfaction of arbitrary goals. It takes into account a model refinement through a finer granularity. The refined goals and obstacles are easy to measure and provide an operational assessment of high-level goals.

## 6. Summary, limitation, and future work

In this paper, we presented an approach for analysing risks in integrating legacy systems with cloud services. The approach is based on goal reasoning. It defines an identify-assess-resolve cycle. The approach uses a probabilistic foundation to formalise goals for adopting cloud services. The probabilities of both, satisfaction and obstacles of goals, are computed in the approach. System architects are able to explore obstacles at the early stage of cloud migration, when there is still flexibility and an affordable resolution still exists. The approach has been applied in scenarios of deploying legacy systems to both Amazon S3 and Microsoft Azure cloud platform.

One limitation of the approach is its reliance on domain information and personal judgment for estimating the probability of occurrence of obstacles for computing goal satisfaction and violation. Accuracy of estimates for leaf obstacles is crucial towards reliability of outcome. In the case of a model with a large number of obstacles, identifying estimates may impose excessive overhead to the decision making process. Continuous feedback from both system developers and end users for the iterative model refinement are an important mitigating factor to alleviate the impact of any subjective judgment.

While we believe that our approach is generic in nature and provides system architects basic modelling techniques and goal/obstacle estimation, we certainly do not claim that the approach validation is conclusive. It is yet to be used in a complete migration project. There are still a few validation steps required to confirm the risk estimates. This will produce more accurate assessment of the produced results. Further comparisons between estimated risks at the early stage of the migration project and subsequent events post migration stage are needed. This will enable us to fully appraise the reliability of the approach. Towards this, we just developed a prototype system, CCER (Cloud Computing Experience Repository) to model goals and obstacles (CCER 2018). CCER is an online repository that provides a single access point of commonly occurring obstacles in adopting cloud services and a concomitant refined set of generic resolution tactics. CCER's user interface consists of interactive forms. It enables a system architect to browse and update obstacles and resolution tactics. In its current version, CCER allows an architect to select cloud adoption goals to identify the list of possible obstacles against those goals. The architect can also hone on obstacles of interest. CCER then proposes tactics required to tackle them.

Due to the interactive nature of the framework, automating reasoning and maintenance of goal models are necessary. This is particularly true in large scale goal modelling. We are extending our prototype with a tool that supports this. The tool will support the computations of probabilities in a consistent manner taking into account the interdependencies among goal and obstacles in model elements.

Pahl, C., et al. (2013). A Comparison of On-Premise to Cloud Migration Approaches. Service-Oriented and Cloud Computing, Springer**:** 212-226.

Pepitone, J. (2011). "Amazon EC2 outage downs Reddit, Quora." Retrieved May **17**: 2011.

Potts, C. (1995). Using schematic scenarios to understand user needs. Proceedings of the 1st conference on Designing interactive systems: processes, practices, methods, & techniques, ACM.

Qiu, M. M., et al. (2013). Systematic Analysis of Public Cloud Service Level Agreements and Related Business Values. Services Computing (SCC), 2013 IEEE International Conference on, IEEE.

Rabetski, P. (2012). "Migration of an on-premise application to the cloud."

Rabetski, P. and G. Schneider (2013). Migration of an On-Premise Application to the Cloud: Experience Report. Service-Oriented and Cloud Computing, Springer**:** 227-241.

Saripalli, P. and B. Walters (2010). Quirc: A quantitative impact and risk assessment framework for cloud security. Cloud Computing (CLOUD), 2010 IEEE 3rd International Conference on, Ieee.

Tran, V., et al. (2011). Application migration to cloud: a taxonomy of critical factors. Proceedings of the 2nd International Workshop on Software Engineering for Cloud Computing, ACM.

Tsidulko, J. (2016). "The 10 Biggest Cloud Outages Of 2016." Available at http://www.crn.com/slide-shows/cloud/300081477/the-10-biggest-cloud-outages-of-2016-so-far.htm.

van Lamsweerde, A. (2004). Goal-oriented requirements enginering: a roundtrip from research to practice [enginering read engineering]. Requirements Engineering Conference, 2004. Proceedings. 12th IEEE International, IEEE.

van Lamsweerde, A. and E. Letier (2000). "Handling obstacles in goal-oriented requirements engineering." Software Engineering, IEEE Transactions on **26**(10): 978-1005.

Van Lamsweerde, A. and E. Letier (2004). From object orientation to goal orientation: A paradigm shift for requirements engineering. Radical Innovations of Software and Systems Engineering in the Future, Springer**:** 325-340.

Yu, E. S. K. and J. Mylopoulos (1994). Understanding "Why" in Software Process Modelling, Analysis, and Design. Proceedings of the 16th international conference on Software engineering. Sorrento, Italy, IEEE Computer Society Press**:** 159-168.

Zardari, S., et al. (2014). "Cloud Adoption: Prioritizing Obstacles and Obstacles Resolution Tactics Using AHP."
**Dr Fahmideh** is a lecturer at the University of Wollongong (UOW). Before this, he has been a postdoctoral researcher at the University of Technology Sydney (UTS). Prior to that, he received a PhD degree in Information Systems from University of New South Wales (UNSW) Sydney. Before joining the academia, Dr Fahmideh has 8 years of industry experience as a software engineer in implementing logical back-end and core computational logic of software systems for different industry sectors including accounting, insurance, and defense. His research focuses on creating new-to-the-world artifacts that help organizations in adopting IT initiatives to tackle problems. His research outcome can be in the form of methodological approaches, conceptual models, decision-making frameworks, and software tools. From a research methodology perspective, he embraces the value of methodological pluralism and utilizes both quantitative approaches (e.g., surveys) and qualitative approaches (e.g., interpretive case study, interview, and domain expert review). To continue on his philosophy of methodological pluralism, he subscribes to apply multi-method studies in design science research. His research outcomes which lie at the intersection of cloud computing, data analytics, IoT, and blockchain technologies have been published in peer-reviewed international venues in information systems and software engineering.

**Professor Ghassan Beydoun** received a degree in Computer Science and a Ph.D. degree in Knowledge Systems from the University of New South Wales. He is currently a Professor of IS and Deputy Head of School of Information, Systems and Modeling at University Technology Sydney. He has authored more than 150 papers for international journals and conferences. He investigates the best uses of ontologies and metamodelling in developing methodologies for distributed intelligent systems. His other research interests include agent based systems applications and knowledge acquisition. Areas where he applies his research include disaster management, smart farming, e-learning, fintech and health informatics.

**Professor Graham Low** received the BE and PhD degrees from The University of Queensland. He is an emeritus professor of information systems in the School of Information Systems, Technology and Management at the University of New South Wales. His research program focuses on the implementation and adoption of new technologies. This can take the form of new/modified approaches/techniques for information systems development such as methodological approaches to agent-oriented information systems design and management of the information systems design and implementation process.